# Computational property predictions of Ta-Nb-Hf-Zr high-entropy alloys

Shashank Mishra[1], Soumyadipta Maiti[1*], Beena Rai[1]

[1]TCS Research, Tata Research Development and Design Center, 54-B Hadapsar Industrial Estate, Hadapsar, Pune, 411013, Maharashtra, India.

*Corresponding author email: soumya.maiti@tcs.com

**ABSTRACT**

Refractory high entropy alloys (R-HEAs) are becoming prominent in recent years because of their properties and uses as high strength and high hardness materials for ambient and high temperature, aerospace and nuclear radiation tolerance applications, orthopedic applications etc. The mechanical properties like yield strength and ductility of TaNbHfZr R-HEA depend on the local nanostructure and chemical ordering, which in term depend on the annealing treatment. In this study we have computationally obtained various properties of the equimolar TaNbHfZr alloy like the role of configurational entropy in the thermodynamic property, rate of evolution of nanostructure morphology in thermally annealed systems, dislocation simulation based quantitative prediction of yield strength, nature of dislocation movement through short range clustering (SRC) and qualitative prediction of ductile to brittle transition behavior. The simulation starts with hybrid Monte Carlo/ Molecular Dynamics (MC/MD) based nanostructure evolution of an initial random solid solution alloy structure with BCC lattice structure created with principal axes along [1 1 1], [-1 1 0] and [-1 -1 2] directions suitable for simulation of ½[1 1 1] edge dislocations. Thermodynamic properties are calculated from the change in enthalpy and configurational entropy, which in term is calculated by next-neighbor bond counting statistics. The MC/MD evolved structures mimic the annealing treatment at 1800°C and the output structures are replicated in periodic directions to make larger 384000 atom structures used for dislocation simulations. Edge dislocations were utilized to obtain and explain for the critically resolved shear stress (CRSS) for the structures with various degrees of nanostructure evolution by annealing, where extra strengthening was observed because of the formations of SRCs. Lastly the MC/MD evolved structures containing dislocations are subjected to a high shear stress beyond CRSS to investigate the stability of the dislocations and the lattice structures to explain the experimentally observed transition from ductile to brittle behavior for the TaNbHfZr R-HEA.

**INTRODUCTION**

High entropy alloys (HEAs) are essentially solid-solution alloys of 4 or more principal elements and getting popular for more than last 15 years since the discovery for their superior structural, mechanical and functional properties[1, 2]. The average structure of the alloys usually comprises of one to two phases of face centered cubic (FCC) or body centered cubic (BCC) lattice[3, 4]. Some of the advanced high performance properties and applications observed and proposed for HEAs are the high strength and ductility at ambient and high temperatures[1, 3, 4], high hardness and wear resistance[4], diffusion barriers in microelectronics[5], binder for cutting tools[6], cryogenic application[7], radiation tolerance[8], enhanced oxidation resistance and hot hardness compared to bearing and high speed steel etc.[9]. The single-phase solid solution stability of near-equimolar concentration of elements in HEAs is attributed to the increased configurational entropy of the systems as $Rln(n)$, where $R$ is the universal gas constant, $n$ is the number of constituent elements in the system. The high-performance properties are mostly attributed down to the locally distorted solid solution lattice structure, sluggish diffusion and lower stacking fault energy of the studied HEAs[1, 4, 5, 7, 8].

Refractory principal elements in HEAs were first used in 2010 by Senkov et al[10] primarily for high temperature (above 1100°C) sustaining high-strength components in aerospace applications. Moreover, other applications of refractory HEAs (R-HEAs) like as high-strength materials at ambient temperatures[11,12], use for electrical resistors, medical implants and micro-electromechanical systems[13,14], radiation tolerance and nuclear applications[15] make them very important and futuristic in nature. It has been observed in the literature that the basic mechanical properties like yield strength, hardness and ductility of single-phase average structure R-HEAs can largely depend on the thermal processing/ annealing treatments for various time and temperature[12,15,16]. In the literature experimental studies on R-HEAs and other complex concentrated alloys (CCAs) have established that the mechanical properties are largely correlated to the change in short-range ordering/ clustering (SRO/ SRC) of various constituent elements, nanostructure phase-instabilities and nanostructure morphologies arising due to the various processing routes[12,15,16,17]. All these processing parameters and related microstructure becomes more relevant in this era of new manufacturing technologies like additive manufacturing (AM) and power metallurgy, where it is observed that an increase in laser power of AM techniques like directed energy deposition or selective laser melting can stabilize single phase average structures of HEA compositions, which is difficult to stabilize otherwise[18]. These types of new manufacturing technologies can accelerate and open up new HEA and R-HEA based materials development fields for aerospace, energy, transportation, biomedical orthopedic, and catalysis applications[10, 12, 18].

However, the related experiments on R-HEAs involved in these literature comprise of atom probe tomography (APT), high-resolution transmission electron microscopy (HRTEM), single-crystal synchrotron beam X-ray diffraction and universal testing machine, which are very costly and involve cumbersome characterization equipment which need highly skilled professionals. The application of computational techniques to predict the properties of R-HEAs could provide an alternative to tedious experiments. In literature there had been some effort to computationally predict some of the properties of TaNbHfZr R-HEAs like the sequential nanostructure evolution, nature of SRO/SRC and local lattice distortions[19]. But the domain size of the simulated R-HEA system was rather small and the hybrid Monte Carlo/ Molecular Dynamics (MC/MD) based thermodynamic properties were only limited to change in enthalpy. There is a lack of comprehensive property prediction study like change in thermodynamic free energy due to high-temperature annealing, changes in nanostructure due to SRO and resulting changes in mechanical properties available from the existing literature. This type of physics-based simulations of microstructure, thermodynamics and resulting properties are becoming important in the emerging field of Integrated Computational Materials Engineering (ICME) in which a large number of experimental trials are reduced and replaced by computational structure-property relationships depending on the processing routes, ultimately aimed at greatly reducing the materials development time and cost[20].

In this study an equimolar TaNbHfZr R-HEA has been computationally studied for the long-term annealing treatment at 1800°C and the resulting effects on change in nanostructure, Gibbs free energy, yield strength and ductility quantitatively or semi-quantitatively. To predict the mechanical strengthening like yield strength, edge dislocations have been simulated for random solid-solution HEAs in recent literature and it has been observed that the edge dislocation can be even better suitable than screw dislocations to predict the strengths of HEAs over a wide temperature range with the help of embedded atom method (EAM) potentials[21]. But the effect of any local short-range order and nano-precipitate evolution due to thermal processing and the resulting large changes in mechanical properties has not been studied and validated in the literature with the help of physics-based models of thermodynamics and dislocation theories. In this study computationally annealed TaNbHfZr R-HEA at high temperature are studied for their equilibrium Gibbs free energy and the corresponding MC/MD generated microstructures are again simulated with edge dislocations for quantitative prediction of yield strength and qualitative study of ductility. The simulations for this study are done with the help of EAM potentials as also used in Ref.[19] for the same TaNbHfZr alloy system.

**RESULTS AND DISCUSSION**

**Thermodynamics**

It is widely accepted that the single phase solid-solution phase stabilization of HEAs is attributed to the increased configurational entropy with the increase in number of elements[1,4,11]. Additionally, there are other types of entropies also associated with alloys such as vibrational, electronic and magnetic entropy. In substitutional solid solution alloys, it has been found that that change in vibrational entropy between ordered and disordered phases is under 0.2 *R*/mol/K, which is around one order of magnitude lower than the configurational entropy of formation[22]. Other entropy sources like electronic and magnetic entropies are expected to be tiny for non-magnetic alloy systems[22]. Thus in this study, focus has been given on the contribution of configurational entropy towards the changes in Gibbs free energy along with changes in enthalpy of the MC evolved systems. In the literature, it has been anticipated that presence of SRO/SRC can significantly change the configurational entropy from the ideal estimate of *R.ln(n)*[22]. Experimental characterizations by atom probe tomography (APT) on 2073 K annealed TaNbHfZr revealed prominent co-clustering of Zr and Hf on {1 0 0} set of habit planes[12]. In this study, the effect of SRC on the overall thermodynamic properties are computationally studied using the quasichemical model found suitable to take into account the SRO into configurational entropy in multi-component alloys[23]. For this, the next-neighbor bond counting statistics for various atomic pair type is applied in the following formula.

$$S = -R . \sum_i x_a \ln(x_a) - \left(\frac{Z}{2}\right) R \left[ \sum_i x_{aa} \ln\left(\frac{x_{aa}}{x_a . x_a}\right) + \sum_j x_{ab} \ln\left(\frac{x_{ab}}{2 . x_a . x_b}\right) \right]$$

Where, $S$ is the entropy, $R$ is the universal gas constant, $x_a$ is the mol fraction of element *a*, $x_{aa}$ is the fraction of next neighbor bonds between the same elements *a*, $x_{ab}$ is the fraction of next neighbor bonds between dissimilar elements *a* and *b*, $Z$ is the coordination factor, respectively[23]. Here the $Z$ is taken as 2, because for a linear solid model that was the value taken and it was found to successfully model a wide varieties of binary and ternary alloy phase-diagram related thermodynamics[23]. The hybrid MC/MD simulation in this study is done in 48000 atoms and the volume of the system corresponds to a domain size of more than 10 nm. The experimentally determined coherent domain size from diffraction studies were found to be also around 10 nm[12] and it is expected that this thermodynamic simulation model should sufficiently capture the enthalpy, entropy and Gibbs free energy changes. Fig. 1 shows the resultant thermodynamic changes in the hybrid MC/MD evolution for annealing treatment at 1800°C using the relation *ΔG = ΔH - T. ΔS*.

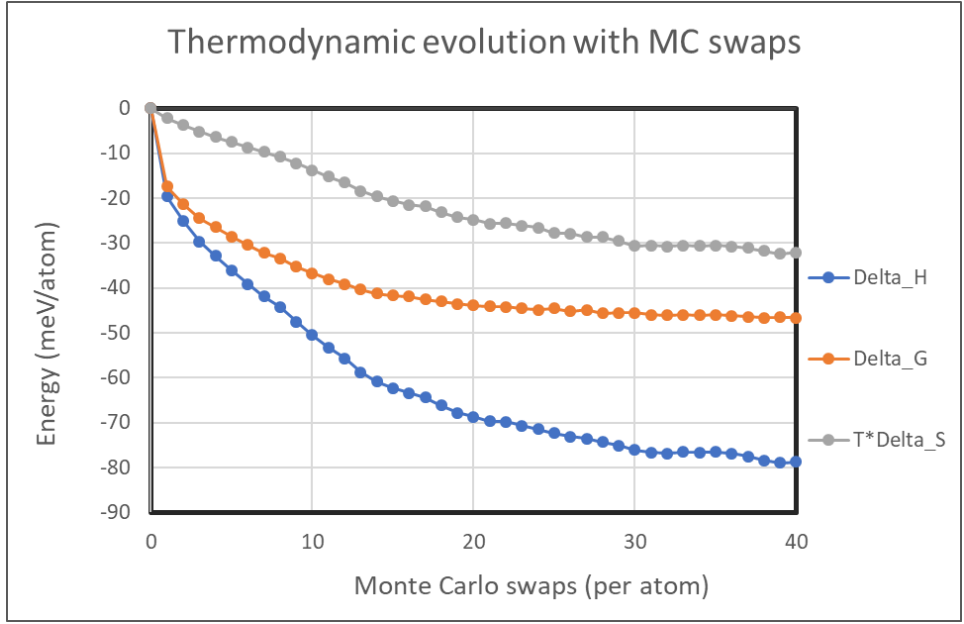

Figure 1. Change in enthalpy, Gibbs free energy and entropic contributions with the MC/MD structure evolution at 1800°C.

It can be observed from Fig. 1 that the resulting $\Delta G$ change is happening continuously and smoothly throughout the whole MC/MD structure evolution. For the initial 5 MC/MD swaps/atom (henceforth called MC swaps), the thermodynamic changes are largely driven by change in enthalpy over entropic contributions. From 5 to 10 MC swaps, both enthalpy and free energy continues to continuously decrease but their mutual difference increase due to increased change in entropy. However, after 10 MC swaps, the rate of change of $\Delta G$ slows down significantly and beyond 20 swaps the $\Delta G$ curve flattens out. The change in $\Delta G$ between 30 and 40 MC swaps converges within 2.5%, which indicates that the thermodynamic equilibrium of the system has been closely reached in this study. At the thermodynamic equilibrium $\Delta H$ is reduced by around 75-80 meV/atom, whereas the $\Delta G$ is reduced by 45 meV/atom. This indicates a significant contribution from change in configurational entropy due to the deviation from the ideal random solid solution behavior due to the local chemical ordering is taking place in the thermodynamics of TaNbHfZr HEA.

**Microstructure**

The Monte Carlo evolved structures created with principal axes along dislocation glide direction X: [111], glide plane normal Y: [-1 1 0] and dislocation line direction [-1 -1 2], commonly used for body centered cubic (BCC) metals, are analyzed[21, 24]. For the better visualization of the SRO/SRC created domain structure, all periodic images of the original 48000 atom structure are also taken. Fig. 2 shows the sequential structure evolution elemental colormap of initial 0 MC swap, 5 MC swaps and until 40 MC swaps at a cross section in the (-1 1 0) plane. The morphology of nanostructure evolution becomes rather relatively unchanged after 20 MC swaps and this is expected from the convergence of $\Delta G$ change of the system discussed earlier. Since the axes of the simulation box are tilted towards the dislocation glide direction, the evolving Zr and Hf rich SRCs appear in angles. However, the angle between [1 1 1] direction and the SRCs make a 55° angle as expected for the SRCs forming in {1 0 0} habit planes as found experimentally[12]. The SRCs gradually form an interconnected grid pattern with is also found experimentally in HRTEM investigations shown is Fig 2g. It can be found that the domains created by the MC structure evolution matches with the experimental morphology having similar domain size of around 10 nm. This grid-type of domain morphology is also found recently in annealed Al-Nb-Ta-V-Ti-Zr based HEA[16]. The MC evolved morphology of annealed TaNbHfZr R-HEA seems to resemble similar "superalloy-like" nanostructure found experimentally in other Al containing R-HEAs[12]. It appears that after 20 MC swaps, since the change in Gibbs free energy is small, there is not much change in the simulated nanostructure which resembles the experimental nanostructure for the 4-days annealed alloy.

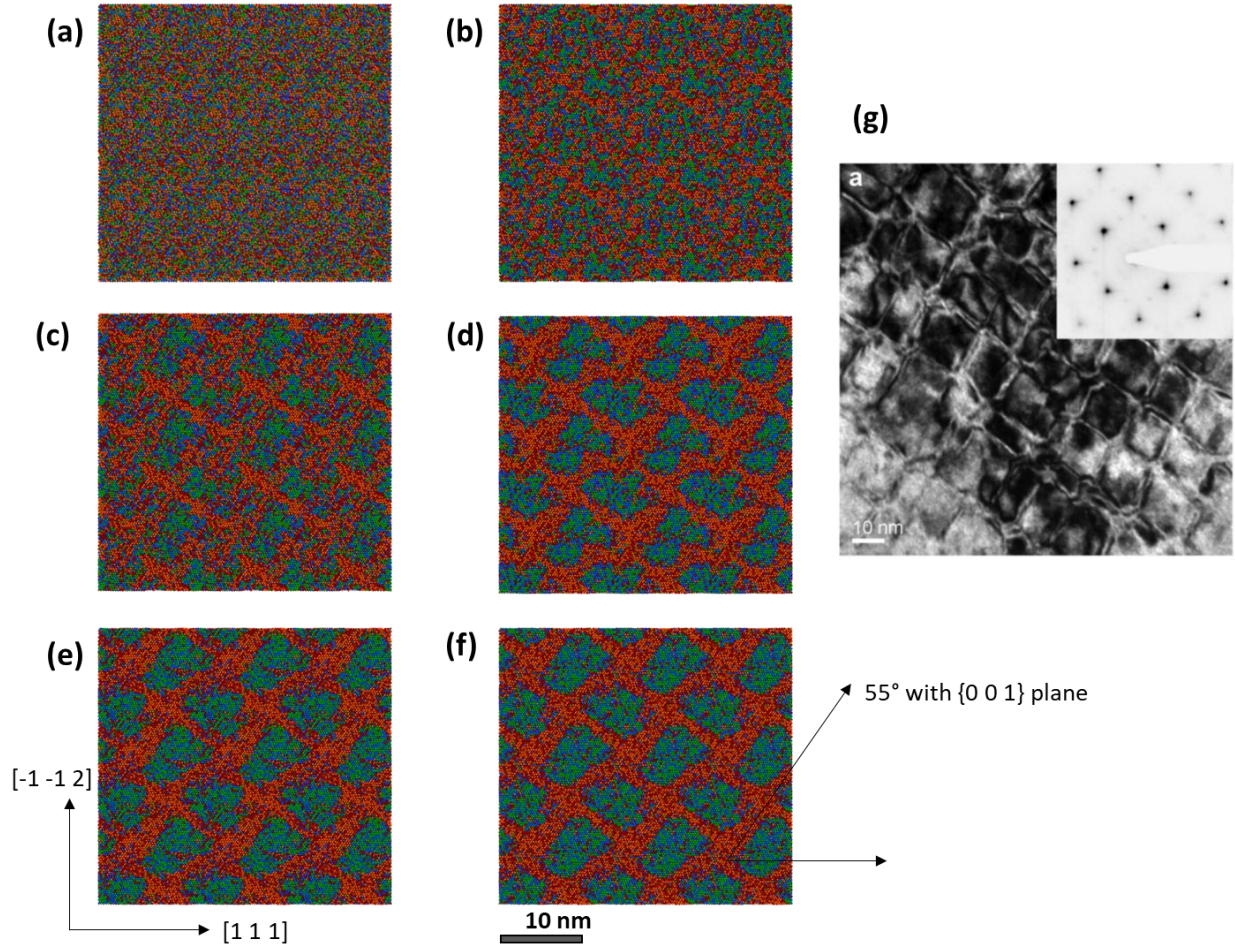

Figure 2. Cross section of MC evolved structure with different colored atom types: (a) 0 swap initial structure, (b) 5 swaps, (c) 10 swaps, (d) 20 swaps, (e) 30 swaps, (f) 40 swaps, respectively. Ta, Nb, Hf and Zr atoms are depicted with green, blue, maroon and orange colors, respectively. Fig. 2g is adapted from Ref.[12]. Fig. 2g is the HRTEM image of the 4-day annealed R-HEA taken along [1 0 0] zone axis with selected area diffraction given in the inset.

**Dislocation and Strengthening**

The edge dislocations introduced by Osetsky method[24] was relaxed for the equilibrium atomic positions so that the dislocation core separates out after CG relaxation. This dislocation insertion is done in the systems starting from initial random solid-solution 0 MC swap structure to 40 MC swap evolved structures. Some of the different evolved structures (0, 5 and 10 MC swaps) containing the edge dislocation along the (-1 10) glide-plane is plotted in Fig. 3 for their various stages of glide movement under the application of the shear stress. Figs 3a1 - 3a3 are related to initial 0 MC swap, Figs. 3b1 – 3b3 are for 5 swaps and Figs. 3c1 – 3c3 are for MC 10 swaps, respectively. In Fig. 3 the blue color atoms depict BCC crystal lattice in the local neighborhood and the white atoms denote deviation from the local BCC lattice, which is created by the common neighbor analysis (CNA) feature in OVITO[25]. The dislocation line is visible with continuous white colored atoms arranged along a curvy line and the core of the dislocations are highlighted by a green line generated by DXA analysis. Under the application of shear stress, the dislocation line starts moving from left to right of the simulation boxes. As the dislocations are created with PAD configuration it exits the simulation box from the right hand side and subsequently enters from the left side due to the periodicity involved.

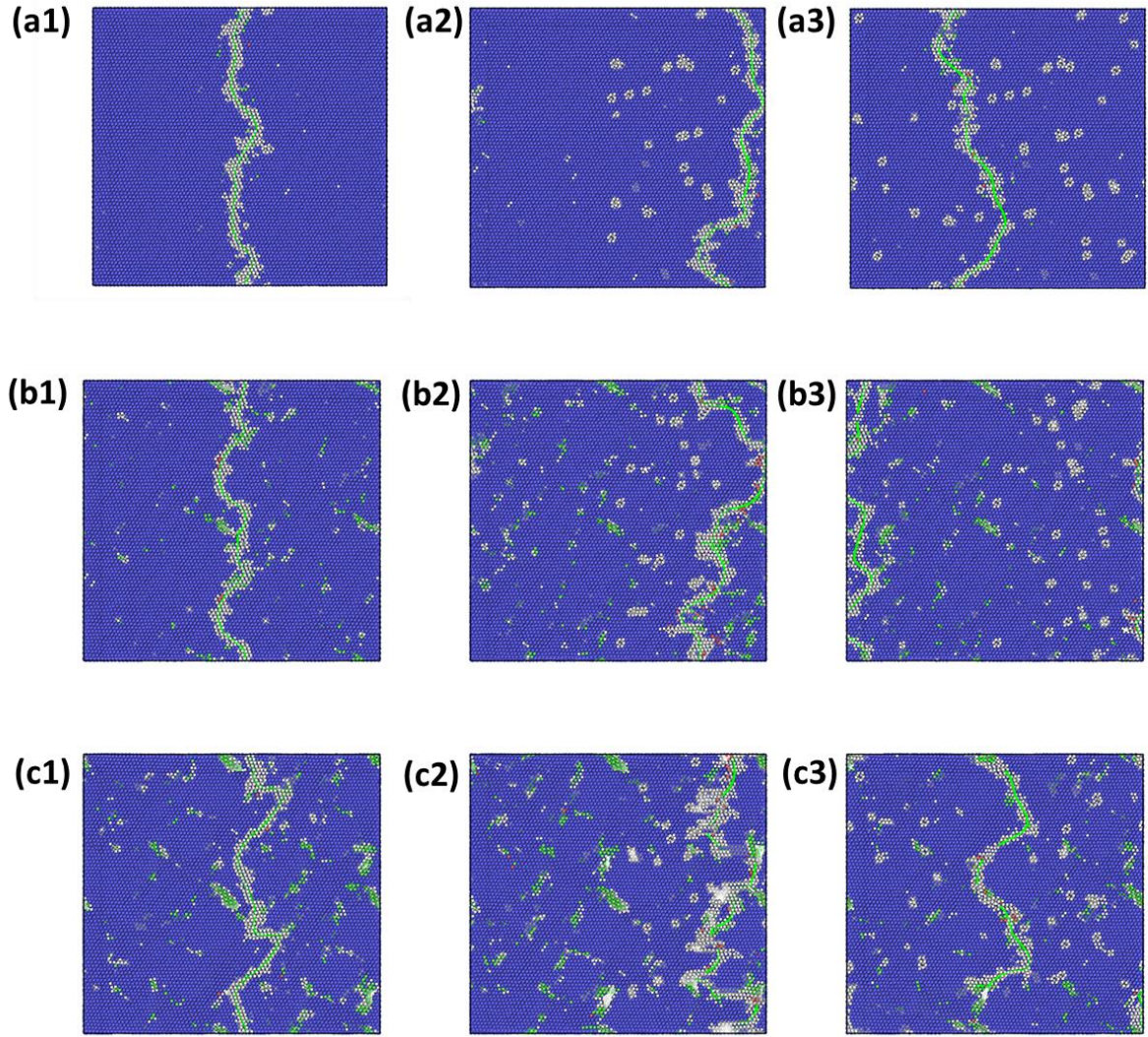

Figure 3. Edge dislocation movement from left to right under applied shear stress: a1-a3 are stages for 0 swap, b1-b3 for 5 swaps and c1-c3 for 10 MC swaps.

It can be seen from the Figs 3a1, 3b1 and 3c1 that when the dislocation is first introduced and relaxed, it takes a zig-zag form with the core of the dislocation having many kinks. This kind of local waviness and kinks of dislocations has also been found in other dislocation-based simulations in the HEAs and multi-element based CCAs[21, 26]. This rough and wavy nature of the core of the dislocation has been attributed to the nanoscale variation of local energy landscape and the ability of the dislocation to find the minimum energy path in that energy landscape[21,26,27]. By the application of shear stress at 300 K to simulate ambient condition behavior, above some threshold value of critically resolved shear stress (CRSS) the dislocation starts to move towards right. But in the process many parts of the dislocation appear to get locally pinned due to various energy minima as also observed in the literature for HEAs[26, 27]. The applied shear stress on the system was gradually applied at different stress intervals and the CRSS level is found where the dislocation was able to overcome the local pinning effects and able to cross the whole length of the system along [1 1 1] direction. At the estimated CRSS, the dislocation moves for around 100Å distance in a simulation time of 50 ps. Similar procedure was followed in the literature for the estimation CRSS by application of constant stress[26,28]. It can be observed from Figs 3a1 to 3a3 that as the edge dislocation moves from center to the right in random R-HEA, it leaves behind atoms other than local BCC lattice atoms. This type of leftover defect atoms are termed as debris and also observed in HEA systems[26]. The waviness of the dislocations even in the random alloy structure (Figs 3a1 to 3a3) and local pinning effects observed in the dislocation motion indicates a high degree of rugged atomic landscape created by the local statistical variation of chemical environments, which the dislocations have to overcome by leaving behind vacancy and interstitial debris[26,27,28]. Similar effects of formation of debris atoms in the dislocation glide path is also observed for MC evolved structures for annealing at 1800°C in Figs. 3b1-3c3. However, the amplitude of local waviness of dislocations in Fig. 3 appears to have increased for the MC evolved annealed systems (5 and 10 swaps), which is expected from even stronger dislocation pinning and obstacle effects from the chemical short-range order.

CRSS values were obtained by studying the edge dislocation movements for the MC evolved structures for every 5 MC swaps and are plotted in Fig. 4. For the initial random alloy with 0 swaps the simulated CRSS was found to be 435 MPa for a total of 381600 atoms containing the dislocation. There is a separate simulation for CRSS done with a much bigger system with 1.29 million atoms and the CRSS for that was also found to be 425 MPa, close for that obtained for the smaller system. This implies that the studied CRSS values of MC evolved structures can be reliable estimates as the simulation boxes have sufficient system sizes so that the sizes don't affect the computed stress levels. This is in accordance with the similar convergence of CRSS values obtained for BCC Fe with system sizes of 80$b$, 120$b$ and 240$b$ respectively, where $b$ is the length of ½[1 1 1] dislocation Burgers vector[24]. The MC evolved structures also have dimensions close to the calculated 80$b$ value of the studied R-HEA. For the initial random solid solution R-HEA, the calculated polycrystalline material yield strength based on CRSS value is given by $\sigma_{ys} = M.\tau =$ 3.067*435 MPa = 1334 MPa, where $M$ is the Taylor factor 3.067[21]. Since there is no chemical order present in the initial random solid solution, any other contributing factor of strengthening may come from the grain size effect of the studied TaNbHfZr. From experimentally studied very similar TaNbHfZrTi alloy, the relationship of grain size related strengthening was found to be $\frac{240}{\sqrt{D}}$ MPa, where $D$ is the grain size in μm[29]. Since the experimentally observed grain sizes of the TaNbHfZr is more than 100 μm[12], the effect of grain size on the yield strength is expected to be less than 25 MPa in polycrystalline material and the strengthening for this R-HEA rather totally arises from solid-solution related strengthening as discussed. Interestingly, the experimentally measured yield stress for as-synthesized TaNbHfZr R-HEA is 1315 MPa, which is only 1.5 % different from the yield strength calculated from CRSS analysis[12]. This kind of ability of the edge dislocation to predict the yield strength is also supported in the literature, where the edge dislocation predicts the strengths of multi element high-concentration BCC alloys even better than the screw dislocation for wide temperature ranges[21]. In conventional BCC single principal element alloys the strengthening is attributed to the screw dislocation behaviors and the kink-pair nucleation process[21, 27]. However, in multi principal element alloys, the movement of dislocations are severely restricted by the locally fluctuating energy and atomic landscape with high degree of activation barrier[21,26,27]. In HEAs the motion of dislocations are rather controlled by the dislocation nanoscale segment detrapping (NSD) from the local energy barriers, and both the edge and screw dislocation movements are governed by the NSD mechanism under the application of stress unlike the mechanism for pure BCC elements[27].

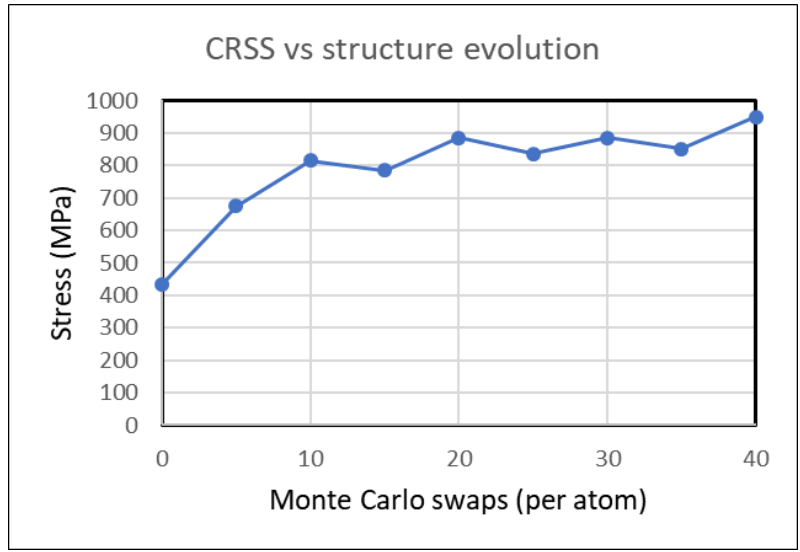

Figure 4. Variation of critically resolved shear stress for TaNbHfZr R-HEA w.r.t. Monte Carlo structure evolution.

Fig. 4 shows that during the first 10 MC swaps of structure evolution, the CRSS value rapidly increases from initial 435 MPa to 815 MPa. This implies that the calculated yield strength reaches a value around 2500 MPa from the Taylor factor relation. The nanostructure morphology after 10-15 Monte Carlo evolution swaps resembles the HRTEM and atom probe tomography based experimental nanostructures of 1-day annealed samples[12]. The experimental yield strength of the R-HEA was found to be 2310 MPa, which is only 7.6% deviating from the value obtained from the simulation techniques[12]. In both the simulated and experimental studies of CRSS and yield strength of this R-HEA, it can be found that the extra strengthening due to chemical SROs happen rapidly within 10 MC swaps for simulation and 1 day of annealing time in experiments. The CRSS values obtained for structures evolved beyond 10 MC swaps remain in the range of 785-885 MPa as similarly the experimentally measured yield strengths remain high above 2000 MPa beyond 1 day of annealing[12].

Interaction of dislocation with the local nanoscale chemical SRCs of the R-HEA revealed direct effect into the strengthening from both experiments and simulations[12]. To investigate more into the nature of dislocation interactions in MC evolved structures, the 10 MC swap system was chosen as this system first reaches the upper range of strengthening corresponding to a 1-day annealed R-HEA. In Fig. 5 perspective images of dislocation core structures along with the crystallographic directions are shown for different stages of dislocation movement under the application of CRSS. The dislocation core structure with green color depicts ½< 1 1 1> type Burgers vector and pink line denotes <1 0 0> type Burgers vectors as extracted by DXA analysis[30]. In Fig 5a the dislocation line in the initial stage looks wavy in nature as in Fig. 3c1. Under the application of CRSS the dislocation bows out, but the motion is restricted by some local pinning points which the dislocation was able to overcome by the discussed detrapping NSD mechanism[26, 27]. Due to periodicity of the PAD configuration involved, the moving bowed out dislocation enter the simulation box from the opposite direction shown in Figs 5b, 5c. In the process the dislocation becomes very wavy and the bowed-out part takes a mixed edge and screw character where edge segments are perpendicular to the Burgers vector and screw segments are parallel to the Burgers vectors. At the stage in Fig. 5c the closely parallel screw segments attract, collapse and touch each other to reduce the length of the dislocation loop and the main dislocation line would pinch-off from the locally closed loop segments. By observing the dislocation movement stages between Figs. 5d-5h it can be found that the dislocation is strongly pinned at the middle section of the simulation box making the original screw dislocation to bow-out and become significantly parallel to the [1 1 1] direction. In the process of overcoming the nanoscale chemical SRC precipitate, the main dislocation forms another small dislocation loop left behind the dislocation SRC interaction site. This small dislocation loop remains at the same place even under the application of stress. This kind of interaction of edge dislocation overcoming a nanoscale obstacle is very similar to the theorized Hirsch mechanism, where also the edge dislocation ultimately crosses the obstacle by leaving a dislocation loop behind the obstacle particle[31].

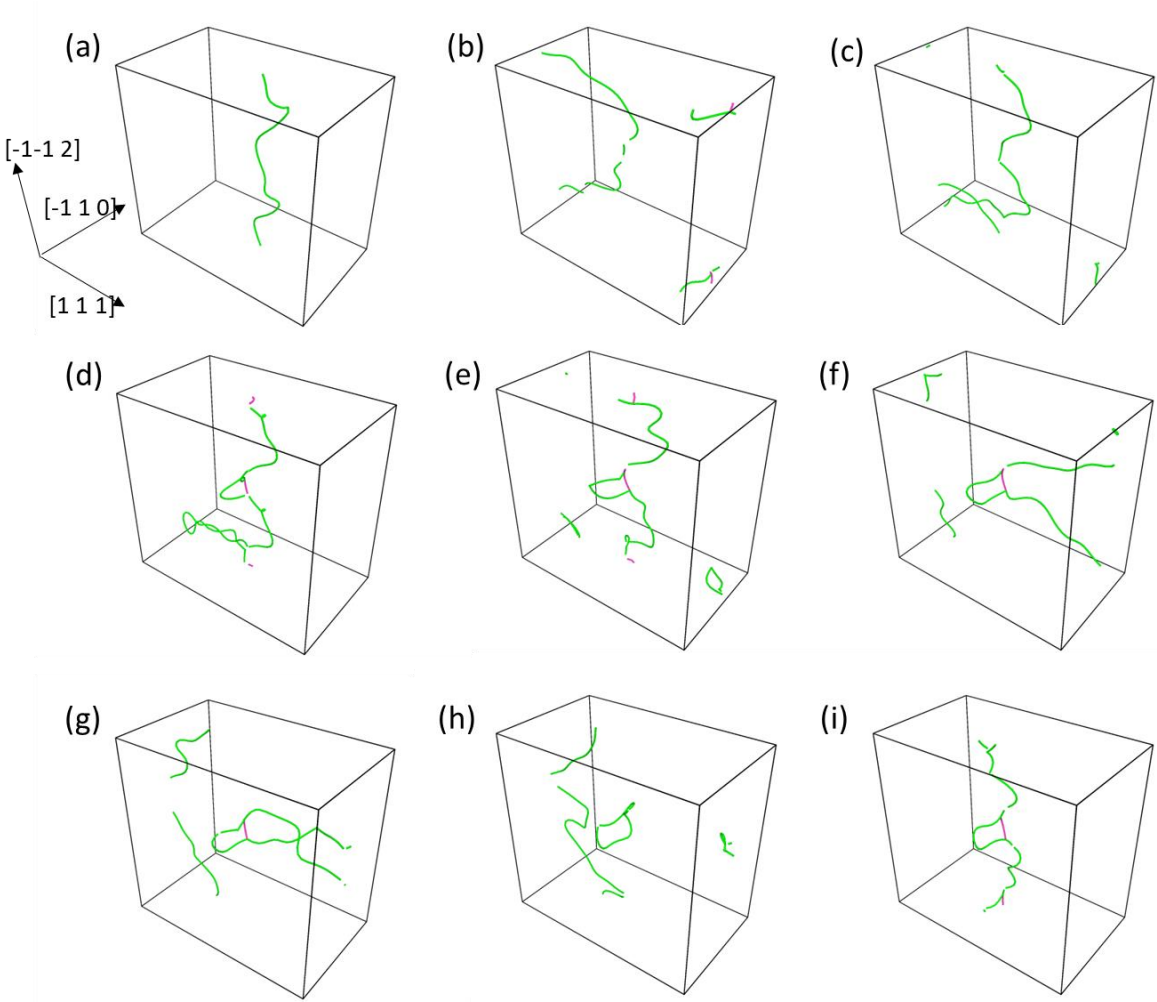

Figure 5. Perspective images of dislocation core movement through the simulation box of 10 MC swap structure under application of CRSS sequentially obtained by DXA method. Green lines denote ½< 1 1 1> family of dislocation Burgers vector and pink line denotes <1 0 0> family of Burgers vectors.

However, in this case the main ½[1 1 1] dislocation appears to have undergone a dislocation reaction while trying to overcome the nanoscale SRC as evidenced by the appearance of the pink colored <1 0 0> type dislocation as seen in Figs 5d-5f from DXA analysis. It was observed from Fig 5e and DXA based Burgers vector analysis that the inserted original ½[1 1 1] dislocation dissociates under applied stress to produce a ½[-1 1 1] dislocation and another [1 0 0] dislocation segment according to the reaction

½[1 1 1] = ½[-1 1 1] + [1 0 0]

Where, the resulting ½[-1 1 1] dislocation is the green color loop and [1 0 0] dislocation is the short pink colored line at the middle part of simulation boxes in Figs. 5d-5f. The ½[-1 1 1] dislocation loop forms out of the main (-1 1 0) glide plane. In Fig. 5f the original inserted ½[1 1 1] edge dislocation line becomes much elongated by the pinning action of the SRC obstacle and the dislocation takes the shape of two parallel screw-type dislocation lines attracting each other because of opposite line sense. In Fig 5g the two attractive bowed-out segments of original ½[1 1 1] attract and touch each other, and then the main ½[1 1 1] segment is pinched-off. In Fig 5g the pinched-off main 1/2[1 1 1] dislocation leaves behind another dislocation loop in the (-1 1 0) glide plane with ½[-1 -1 -1] Burgers vector, because of the opposite line sense of this newly created loop. However, this ½[-1 -1 -1] dislocation loop shrinks and spontaneously reacts with the previously created [1 0 0] according to the reaction

½[-1 -1 -1] + [1 0 0] = ½[1 -1 -1]

where ½[1 -1 -1] is the line segment created in the place of the pink colored [1 0 0] segment. This in term forms and completes the green colored loop of ½<1 1 1> dislocation type observed in Fig. 5h. Subsequently, the main ½[1 1 1] dislocation enters from the opposite end due to the PAD configuration and reacts with the green 1/2<1 1 1> loop created by the previous dislocation reactions. The reaction product is shown in Fig. 5i, where we again find the occurrence of the pink colored short [1 0 0] dislocation segment. This is explained by the following reaction

½[1 1 1] + ½[1 -1 -1] = [1 0 0]

where ½[1 1 1] is the main dislocation entering into the system and [1 0 0] is the created pink colored segment in Fig. 5i. Interestingly, this type dislocation configurations in Fig 5i is similar to the configurations in Figs. 5d-5f with respect to the types of dislocations present, their orientation and all. So the configurations shown in Figs 5e to 5i are periodic with time in nature and the subsequently entering dislocations experience the effect of repeated dislocation reactions happening at the increased CRSS of the 10 MC swap system. In the DXA analysis done with the other higher MC swap configurations than 10 swaps, we see the similar types of dislocation interactions and CRSS values that we already observe in the 10 MC swap system. However, in the initial random alloy and the alloys with lower degree of chemical ordering in 5 MC swap system, the original inserted ½[1 1 1] dislocation line remains intact due to lower amount of CRSS required and does not dissociate by any dislocation reaction as discussed for 10 MC swap systems. In the literature concerning ½<1 1 1> dislocations of BCC metals, these kinds of back and forth interactions of ½<1 1 1> with <1 0 0> dislocations have been discussed with dislocation multi-junctions, dissociation or unzipping of dislocations under applied force[32]. The <1 0 0> type of dislocation as discussed in literature is observed in dislocation networks of BCC metals and the high expected CRSS of this <1 0 0> dislocation makes them unlikely to move at ambient temperature, which is observed in the above dislocation analysis of the 10 MC swap structures[33].

Experimental investigations of this studied TaNbHfZr system reveals that the compressive ductility at ambient temperature was around 22% engineering strain for the as synthesized random solid solution alloy, whereas the ductility drastically drops below 1% for the 1-day annealed material[12]. This inability to plastically deform for the 1-day annealed material is studied by dislocations here. For any significant plastic deformation to happen, the grain boundary of a crystal has to physically deform the adjacent grain and activate the dislocation emitting source in the next grain[33]. This phenomenon is supposed to be done by the edge dislocation near to the grain boundary which is created by dislocation pile up and local high stress acting on the leading dislocation. The stress acting on the leading dislocation near the grain boundary is given by

$$\tau_1 = n\tau = \frac{\pi L \tau^2}{Gb}$$

where $\tau_1$ is the stress acting on the leading dislocation of the pile-up, $n$ is the number of dislocations in the pile-up, $L$ is the length of the grain, $G$ is shear modulus, $b$ is burgers vector and $\tau$ is the externally applied shear stress[33]. As suggested by the above relation, in a large grain the stress on the leading dislocations increases rapidly with the applied stress. Also, for the required ductility of the material it becomes crucial for the dislocations emitted from a crack tip to quickly nucleate and travel at a faster rate away from the crack[33,34]. This in term increases the shielding effect from the dislocation-based crack closure and prevents the material from behaving brittle[34]. Since the local state of shear stress at the leading dislocations in a dislocation pile-up and near to a crack tip increases rapidly, it is of importance to study the motion and stability of dislocation under high shear-stress conditions for the studied R-HEA. In this study some of the MC evolved structures used for determining CRSS is also subjected to a high shear stress of 1500 MPa. In Figs. 6a1-6a3, the common neighbor analysis (CNA) figures for 5 MC swap system and in Figs. 6b1-6b3 that of 15 swaps system under 1500 MPa shear stress is given. The configurations are taken at the beginning of stress application and then 50 ps simulation time apart. The blue colored atoms are the BCC lattice atoms and white colored atoms are the other amorphous type as obtained from CNA. It can be seen from the perspective Figs. 6a1-6a3 that for the 5 MC swap system, the simulation box remains BCC in the lattice framework throughout the simulation. The dislocation line from DXA analysis was also found to be continuously moving in its PAD configuration with velocity around 1000 m/s. Whereas, in the 15 MC swap system, under the application of the 1500 MPa of shear stress, the initially created dislocation line completely breaks down and could not be traced. Then in the subsequent timesteps, this system undergoes local internal shearing in the interior as evidenced by the accumulation of amorphous white atoms along the glide plane in Fig. 5b2. With the progression of simulation time, this amorphous shear band increases in thickness and the crystalline BCC lattice structure of the simulation box collapses as seen in Fig. 5b3. This type of local breakdown of lattice framework will prevent any further dislocations to be produced and gliding through the material. As this is the situation experienced by the dislocations causing yielding and plastic deformation in dislocation pile-up and crack tip propagation, the high stress simulations indicate a low degree of expected ductility for the 15 MC swap structure. This is getting reflected in experiment on the very low ductility of below 1% for the 1-day annealed alloy, which in term resembles the 15 MC swap evolved structure of this study and the previous study on structure evolution[19].

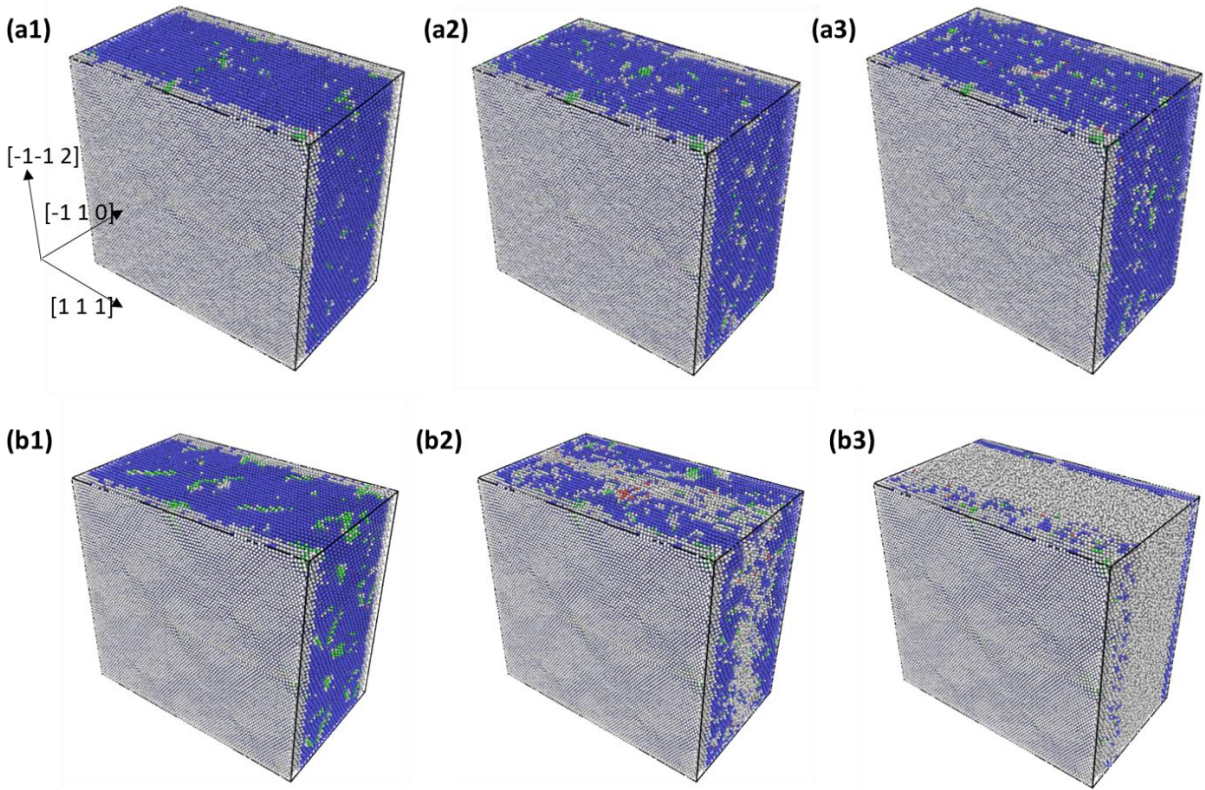

Figure 6. Common neighbor analysis colormap for BCC alloy structure: a1-a3 are for the 5 MC swap and b1-b3 are for the 15 MC swap structure under the application of 1500 MPa shear stress. Structures are taken approximately 50 ps apart in simulation time.

## CONCLUSION

It can be inferred from the whole computational study of TaNbHfZr R-HEA that many of the thermodynamic, microstructural, strengthening, ductility and dislocation related properties can be obtained from the simulations alone. The thermodynamic changes from the initial random solid solution alloy by hybrid MC/MD structure evolution technique indicates that the entropic contribution to the free energy can be almost 40% of the change of enthalpy due to the presence of local chemical clustering. The quasichemical approximation model deployed here can reflect a gradual and smooth change in Gibbs free energy reaching its equilibrium minimum. Due to the MC/MD structure evolution resembling an annealing treatment, the Hf and Zr-rich SRCs form in the expected {1 0 0} habit planes appearing close to the experimental nanoscale superalloy-like morphology even though the structure is made along dislocation glide direction. The dislocation core is wavy in nature due to the rugged local energy landscape in the multi principal element alloy. The amplitude of the edge dislocation waviness increases in the MC evolved configurations due to strong interactions with SRCs to minimize the energy. The dislocation moves under applied shear stress with many local pinning and nanoscale detrapping mechanism leaving behind debris of vacancy or interstitials. The CRSS values obtained from the edge dislocation simulations indicate that the simulated yield strength of the alloy is within 1.5 % accuracy of the experimental value for the random solid solution and within 7.6% of the experimental value for the 1-day annealed material. From the 10 MC swap structure onwards the ½[1 1 1] dislocation undergoes a Hirsch-like mechanism for overcoming the nanoscale SRC obstacles by significantly increasing the CRSS values. The initially inserted ½<1 1 1> dislocation undergoes disassociation reaction to form other ½<1 1 1> and stationary <1 0 0> dislocations in 10 swap structures. There are some back and forth cyclic dislocation reactions involved for the dislocation multijunction networks. Under application of high level of CRSS like 1500 MPa for the 15 MC swap structure, the dislocation was completely broken and destroyed turning the crystalline structure into a amorphous configuration with a shear band. This is in line with the experimentally observed very low ductility of the 1-day annealed alloy, because of the inability of the dislocations to freely pass through the local structure under high shear stress as expected for dislocation pile-ups and crack-tip shielding for ductile materials. The simulation methodology developed and explained in this study can be applied to a wide variety of high-entropy alloys and other complex concentrated alloys to predict their basic mechanical and thermodynamic properties depending on the processing routes. This methodology can accelerate new alloy materials development work by potentially reducing the dependency on a large number of costly and tedious experiments on nanostructure characterization, material synthesis, processing and mechanical testing time and related man hours involved.

**METHODS**

Initially, a system containing 48000 atoms of TaNbHfZr alloy was created along the three orthogonal axes along X = [1 1 1], Y = [-1 1 0] and Z= [-1 -1 2] directions and with this the supercell was created as 20x15x20 by using the package LAMMPS[35]. The atomic fractions in initial random solid solution alloy for the study for Ta, Nb, Hf and Zr were 25.16%, 25.05%, 24.86%, 24.92%, respectively. The lattice parameter (LP) used to create the structure was kept at 3.54 Å as this was found as the equilibrium LP by NPT simulation at 2073 K[19]. Then this structure was subjected to hybrid MC/MD simulation at 2073 K using the methodology given in Ref.[19] until 40 attempted swaps/atom. In the MC/MD procedure two dissimilar atom types were randomly chosen and their positions are swapped. The resulting new configuration is minimized for energy by conjugate gradient method and the new trial configuration is accepted or rejected according to Metropolis Monte Carlo criterion by comparing the potential energies of trial configurations as discussed in detail in Ref.[19]. The systems created after each 48000 attempted swaps were stored. Based on coordinates of the stored MC/MD configurations, next neighbor bond counting statistics of all atomic pair types of the structures are also stored separately. This was useful in calculating configurational entropy and thermodynamic properties as discussed in following sections.

For the estimation of critically resolved shear stress (CRSS), the systems generated after the MC/MD structure evolution steps were replicated by 2x2x2 in each direction periodically in order to get a bigger supercell with 384000 atoms. Later an ½ [111] edge dislocation was introduced in the middle of the MC/MD evolved systems with line direction along Z [-1 -1 2] and glide direction along X [111] using the Osetsky method[24]. However, since the CRSS simulations are compared with experimentally obtained strengthening behavior of the R-HEA at ambient temperature[12], the LP of the 384000-atom structure was adjusted to the 3.49 Å, which was obtained by NPT equilibration for 100 ps. This bigger supercell with dislocation had dimensions around 242 Å, 148 Å and 228 Å along X, Y and Z directions, respectively. Then, these systems were subjected to energy minimization with periodic boundary conditions along X and Z and free boundary condition along Y direction. Further, these systems were equilibrated at 300 K using the NVT ensemble for 100 ps. Then, a pure shear stress was applied on the (-110) glide plane to the top 15 Å of atoms keeping the bottom 15 Å of atoms fixed. Periodic boundary conditions were applied along X and Z directions, whereas free boundary condition was applied along Y direction. This created a periodic array of dislocation (PAD) configuration of simulation[24]. The dislocation simulations were run for ambient temperature conditions by keeping the temperature fixed at 300 K and using NVE microcanonical ensemble simulation in LAMMPS. The motion of the dislocation along with their interaction with SRO atoms were observed. The shear stress level for the next run was adjusted to get the CRSS, which was taken as stress at which the dislocation started to move for around 100 Å in 50 ps and below which it didn't move as much. For all the visualization purpose, software OVITO[25] and for running the MD simulations LAMMPS was used. For identifying the dislocations in the systems, the common neighbor analysis (CNA) and dislocation analysis tools of OVITO were employed. The dislocations were extracted by using DXA algorithm[30] and its movements are subsequently observed under the application of shear stress. The EAM potential used for all the simulations were developed based on physical input parameters of atoms like LP, cohesive energy, vacancy formation energy, elastic constants and its details are given in Ref.[12].

## DATA AVAILABILITY

The datasets generated during and/or analysed during the current study are available from the corresponding author on reasonable request.

## ACKNOWLEDGEMENTS

The authors would like to thank High Performance Computing Group at Tata Consultancy Services (TCS) for providing access to the Gold Server. They would also thank Mr. K Ananth Krishanan, CTO, TCS for his encouragements and support for this project.

**AUTHOR CONTRIBUTIONS**

S. Maiti ideated, designed the study and analyzed the results. S. Mishra carried out the MC/MD structure evolution analysis, dislocation visualization analysis and making the figures. B. Rai contributed to the overall guidance for the work.

**COMPETING INTERESTS**

The authors declare that there are no competing interests.